\documentclass{aa}

\usepackage{graphicx}
\usepackage{amsmath}
\usepackage[version=4]{mhchem}  
\usepackage{txfonts}
\usepackage{siunitx}
\usepackage{xcolor}
\begin{document}

   \title{Negative polarization properties of regolith simulants}

   \subtitle{Systematic experimental evaluation of composition effects}

   \author{S. Spadaccia,
          \inst{1}
          C.H.L. Patty,
          \inst{1}
          H.L. Capelo,
          \inst{1}
          N. Thomas,
          \inst{1}
          \and
          A. Pommerol\inst{1}
          }

   \institute{Space Research and Planetary Sciences Division, Physikalisches Institut, University of Bern, Sidlerstrasse 5, 3012 Bern, Switzerland\\
              \email{stefano.spadaccia@unibe.ch}
              }

   \date{Received date; accepted date}

 
  \abstract
   {Polarization phase curves of asteroids and other small airless bodies are influenced by the compositional and physical properties of their regolith. The mixing of minerals composing the regolith influences the negative polarization at small phase angles because it changes the multiple scattering properties of the medium.}
   {This work aims to demonstrate experimentally how the mixing effect influences the polarization phase curve at small phase angles for different mineralogies relevant for asteroids, and to determine how different aggregate sizes affect the negative polarization.}
   {We prepared a set of binary and ternary mixtures with different common minerals on asteroids and one set of the same mixture with different aggregate sizes. We measured their reflected light at 530 nm with full Stokes polarimetry at phase angles ranging from \ang{0.8} to \ang{30}.}
   {The mixing effect of the mixtures with both bright and dark minerals significantly changes the behavior of the phase curves in terms of minimum polarization, phase angle of the minimum, and inversion angle with respect to the mineral components that are mixed together. The changes in phase curve could explain the polarization observation of particular classes of asteroids (F and L class) and other asteroids with peculiar polarization curves or photometric properties. Furthermore, we demonstrate that the negative polarization is invariant to the presence of dust aggregates up to centimeter sizes.}
   {}


   \maketitle
%

\section{Introduction}
Polarimetry is a powerful tool for studying the properties of many objects in our Solar System and beyond. The induced linear polarization in the light reflected from a surface can provide valuable information about the porosity, multiple scattering, shape of the grains, and their indices of refraction. The challenge lies in disentangling this intricately interwoven mass of information when the polarization of astronomical objects is measured. 

One of the most commonly used methods is the analysis of the relation between the linear polarization $P$ and the phase angle ($\alpha$, i.e., the angle between the reflected light and the light source in the scattering plane). $P$ is given by 
\begin{equation}
    P=\frac{I_{\perp}-I_{\parallel}}{I_{\perp}+I_{\parallel}},
        \label{eq:errD}
\end{equation}
where $I_{\perp}$ and $I_{\parallel}$ are the intensities of the reflected light with the polarization axis normal and parallel to the plane of scattering, respectively. We note that $P=Q/I$ using the Stokes formalism. 

Irregular particles of many bodies in the Solar System show similar features in their polarization phase curve. At very small $\alpha$ ($< $\ang{3}), the coherent backscattering opposition effect (CBOE) occurs and the reflected light increases nonlinearly \citep{shkuratov1989new, muinonen1992light, hapke_theory_1993}. The CBOE causes a surge in reflected circular polarization and a decrease in reflected linear polarization when the object is illuminated with circular and linear polarized light, respectively \citep{nelson_phase_1998}. For unpolarized incident light, the linear polirization at small $\alpha$ (i.e., smaller than $15-$\ang{25}) is negative and thus is in the backscattering regime $I_{\perp}<I_{\parallel}$. The part of the phase curve that is dominated by negative polarization is commonly referred to as the negative polarization branch, and the minimum of polarization $|P_{min}|$  usually falls at phase angles $\alpha_{min}=8-$\ang{15}. The phase angle at which the polarization is zero is called inversion angle $\alpha_{inv}$. At higher $\alpha$ (between 30 and $90-$\ang{100}), the polarization increases and reaches a maximum. Generally, the maximum linear polarization is inversely correlated to the albedo of the surface by the so-called Umov effect \citep{umov_chromatische_1905, zubko_interpretation_2011}. 

The negative polarization feature of asteroids, comets, and other bodies has been studied extensively both theoretically and experimentally, in order to understand its origin and its possible use as a proxy for retrieving small-particle properties through remote sensing (see \citet{levasseur-regourd_asteroids_2015} and \citet{levasseur-regourd_cometary_2018} for a complete review). Within this framework, the relation between different parameters that control the shape of the phase curve has been analyzed. 

Minor bodies such as comets and asteroids are covered by regolith, that is, by a loose deposit of fine dust and rock pieces developed by space weather and meteroid impacts. The shape of the polarization phase curve of dust depends on the number of interactions the incident radiation has with the sample, which can be one time (single scattering) or multiple times (multiple scattering). Single scattering predominantly occurs in environments with low particle density, for example, cometary comae, while regolith-like surfaces are characterized by multiple scattering. \citet{SHKURATOV2004267} and \citet{zubko_interpretation_2011} showed that the polarization phase curves of multiple and single-scattering environments are not identical. Multiple-scattering events effectively scramble the overall polarization signal that is reflected back to the observer, resulting in an overall weaker polarization. The single-scattering polarization phase curves are also usually characterized by higher inversion angles, higher $\alpha$ at the polarization minimum, and higher maximum polarization. 

Many studies have been carried out with the aim to understand the origin of the negative polarization in the case of single scattering.
There are indications that the negative polarization arises from the coherent backscattering effect of the particles \citep{zubko_light_2008}. The single-scattering negative polarization of dust grains is stronger for small particles (below \SI{3}{\micro\metre}), and it also depends on their absorption properties \citep{zubko_light_2013-1,zubko_small_2020-1}. Interestingly, the negative polarization tends to disappear when the submicrometer particles are removed from dust simulant samples \citep{escobar-cerezo_experimental_2018} and from clouds of silicates \citep{munoz_retrieving_2021}. In the case of multiple scattering, the negative polarization is also dependent on the porosity of the sample, showing an increase in amplitude of the negative polarization with sample compression (thus decreasing the surface porosity) and exhibits a dependence on changing the incidence and emission angle after the phase angle is fixed \citep{shkuratov_opposition_2002,halder_dependence_2018}.

Various authors have correlated the presence of particular minerals in asteroid regolith (identified by scalar spectroscopy) to their multiple-scattering polarization properties. \citet{cellino_2016} investigated the variation in negative polarization of asteroid (4) Vesta depending on its rotation and found a good correlation between the surface albedo variations and the polarizance. The authors, however, pointed out that a complete explanation of the polarization data needs to take variations in surface geometry and mineralogical composition into consideration. Particularly, they demonstrated that dark regions dominated by eucrite seem to show higher $|P_{min}|$. More recently, \citet{castro2022polarimetric} observed a variation in polarized light with the rotational curve of asteroid (16) Psyche, correlating with changes in albedo and in surface geometry. \citet{gil-hutton_polarimetric_2017} used the phase curve model of \citet{Muinonen2002} to calculate the refractive index of the regolith on the surface of 129 asteroids, and observed that there is a strong correlation between the refraction index and the inversion angle, and between $|P_{min}|$ and the distance between single scattering particles. A similar result was obtained by \citet{masiero_effect_2009}, who found that the refractive index plays a more important role in determining the inversion angle than the particle size.

It has been demonstrated that asteroids cluster together in the $P_{min}-\alpha_{inv}$ space depending on their class types \citep{belskaya2017refining}. Generally, this indicates that asteroids in the same family share similar mineralogical compositions and physical properties. To some extent, this is also affected by the asteroid albedo. Asteroids with a high albedo in the V band (more than 0.2) have a higher $P_{min}$ than moderate-albedo asteroids (0.1-0.2 in V), and the darkest asteroids (C, Ch, and B classes; an albedo  lower than 0.1) populate the lowest $P_{min}$ region. In this context, \citet{cellino_2016} noted that there is some degree of mixing between moderate- and low-albedo asteroids defined in the region $P_{min}=-1.1\%-1.4\%$ and $\alpha_{inv}=18-$\ang{21} in Fig.~\ref{fig:painv}. \citet{dollfus1989photopolarimetry} interpreted the fact that terrestrial rocks and lunar fines show a smaller and larger polarimetric inversion angle, respectively, than most asteroids as an indication that the surface of asteroids contains coarser material than the surface of the Moon. More recently, two new classes of rare asteroids have been identified \citep{belskaya_f-type_2005,cellino2006strange}: the F-class asteroids, which show small inversion angles ($14-$\ang{16}), and L-type asteroids (''Barbarians``), with inversion angles in the range $\alpha_{inv}=25-$\ang{30}. These asteroid classes are outliers with respect to the normal distribution of asteroids in the $P_{min}-\alpha_{inv}$ space.

It has been suggested that the high inversion angle of L-type asteroids is due to the presence of white spinel-bearing CAIs on the surface \citep{devogele_new_2018,sunshine_ancient_2008}, mixed in a dark matrix \citep{burbine1992s}. Nevertheless, asteroids and their corresponding meteorite classes show a wide range of mineral compositions, including olivine, pyroxene, plagioclase, spinel, and phyllosilicates (\citet{michel2015asteroids} and references therein). While it is possible to directly observe the presence of multiple minerals on a surface through scalar spectroscopy, the effect on the polarization phase curve of such mixtures is not clear. Studies to clarify this were made by \citet{boehnhardt_surface_2004} and \citet{bagnulo_exploring_2006}, who successfully modeled the phase curve of trans-Neptunian objects at very small phase angles, under the assumption of a two-component surface medium composed of bright (ice) and dark particles. 

\citet{shkuratov1987negative} and \citet{shkuratov1994critical} demonstrated that a material with a mixture of small and large albedos can show a different negative polarization phase curve and a higher $|P_{min}|$ than the endmembers of the mixture. This effect was also used by \citet{belskaya_f-type_2005} in order to explain the very small inversion angle of F-type asteroids, which are thought to share some physical properties with comets \citep{cellino2018unusual}. \citet{belskaya_f-type_2005} proposed that because F-class asteroids are very dark (0.03-0.07 albedo) but have a higher $P_{min}$ than C-class asteroids, the surface regolith must be homogeneous, because any mixing effect would decrease $|P_{min}|$. A systematic study of the mixing effect, however, is lacking in the literature. 

The surface roughness can also influence the linear polarizance. The irregular shape (and/or macroscopic roughness) of an asteroid can result in nonzero polarization at $\alpha=0$, which generally results from polarization contributions arising from different parts of the asteroids with a distribution of incidence and emission directions. The Hayabusa spacecraft observed a great variety of surface morphology on (25143) Itokawa, ranging from boulders and rough terrain to flatter terrain characterized by millimeter (mm) to centimeter (cm) gravel. 

While the effect of particle size on the negative polarization at small phase angles is generally understood, there are still important effects that could influence this part of the polarization phase curve, such as the mixing of materials with different optical properties and the aggregation of small particles. The aim of this work is to systematically explore the change  in  polarization phase curve of relevant regolith-like minerals when the minerals are intimately mixed together in different compositions. In addition, we demonstrate the change in polarizance when the powder grains are incorporated into mm-cm size aggregates.

The remainder of this manuscript is structured as follows: in section 2 we describe our method and experimental apparatus, in section 3 we present our results and summarize the findings, and section 4 provides a discussion of the results in the context of asteroid surface features. We conclude in section 5. 

\section{Materials and methods}\label{methods}
\subsection{Experimental setup}\label{setup}

\begin{figure}
\centering
\includegraphics{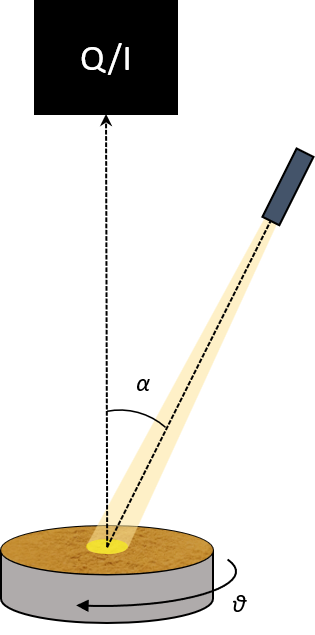}
\caption{Geometry of the POLICES setup. A nearly collimated light source illuminates the sample with a phase angle $\alpha$ that varies between 0.8 and \ang{30}. The sample can be turned in steps of \ang{45} on the azimuth angle $\theta$. The full Stokes polarimeter is is fixed perpendicular to the sample and measures $Q/I$ at different sample angles $\alpha$ and $\theta$.}
    \label{fig:setup}
\end{figure}

The polarization measurements were carried out with the POLarimeter for ICE Samples (POLICES) at University of Bern (see also \citet{Poch_2018} and \citet{patty2022} in revision). POLICES is a full Stokes polarimeter (Dual PEM II/FS42-47, Hinds Instruments) that allows us to measure the polarization state of the light scattered by a sample at different phase angles. It consists of an enclosure in which an arm holding a collimated light source illuminates a sample placed at the bottom of the enclosure. The scattered light is measured at the top of the enclosure in nadir direction. In this configuration, the reflection angle is approximately \ang{0}, while the incident angle, which is thus similar to the phase angle $\alpha$, can be varied. 

The light source is a 530 nm LED (Thorlabs M530F2) that is depolarized and fiber-fed to a collimating head, providing an illuminated sample area with a diameter of approximately 15 mm. The remnant polarization of the incidence light at 530 nm is lower than 0.01\%, which agrees well with the polarization of the global light coming from the solar disk \citep{clarke1996sun}. The arm can span a wide range of phase angles (the angle formed by the incidence light and the emission direction), from \ang{-30} to \ang{75}. The minimum rotation step of the arm is \ang{0.1}. The sample is placed on the same plane that contains the rotation axis of the arm, and it sits on a rotation stage that can change the azimuth of the sample $\theta$ from \ang{0} to \ang{360} (Fig. \ref{fig:setup}). The polarimeter entrance pupil is approximately 50 cm away from the sample, and the fiber used to illuminate the sample is at about 44 cm from the sample. The width of the illuminated spot on the sample has a negligible effect on the estimation of the phase angle (about \ang{0.1}).

\subsection{Samples}\label{samples}
\begin{figure*}
\centering
\includegraphics[width=16.4cm,clip]{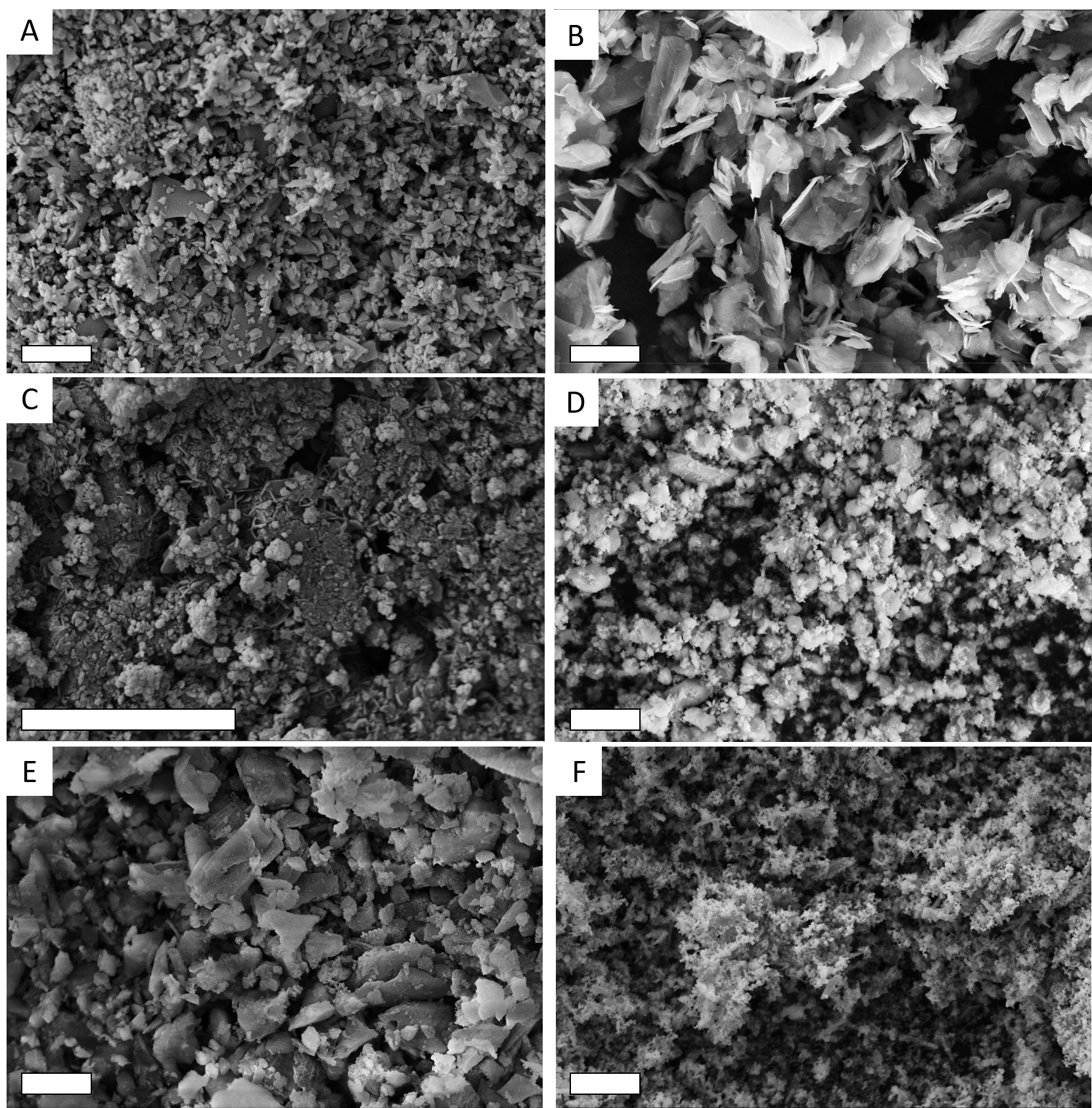}
\caption{Scanning electron microscope images of the six minerals. A) Mg-spinel, B) graphite, C) forsterite, D) silica, E) fayalite, and F) magnetite. All the scale bars represent \SI{10}{\micro\metre}.}
    \label{fig:SEM}
\end{figure*}

\begin{figure*}
\centering
\includegraphics[width=16.4cm,clip]{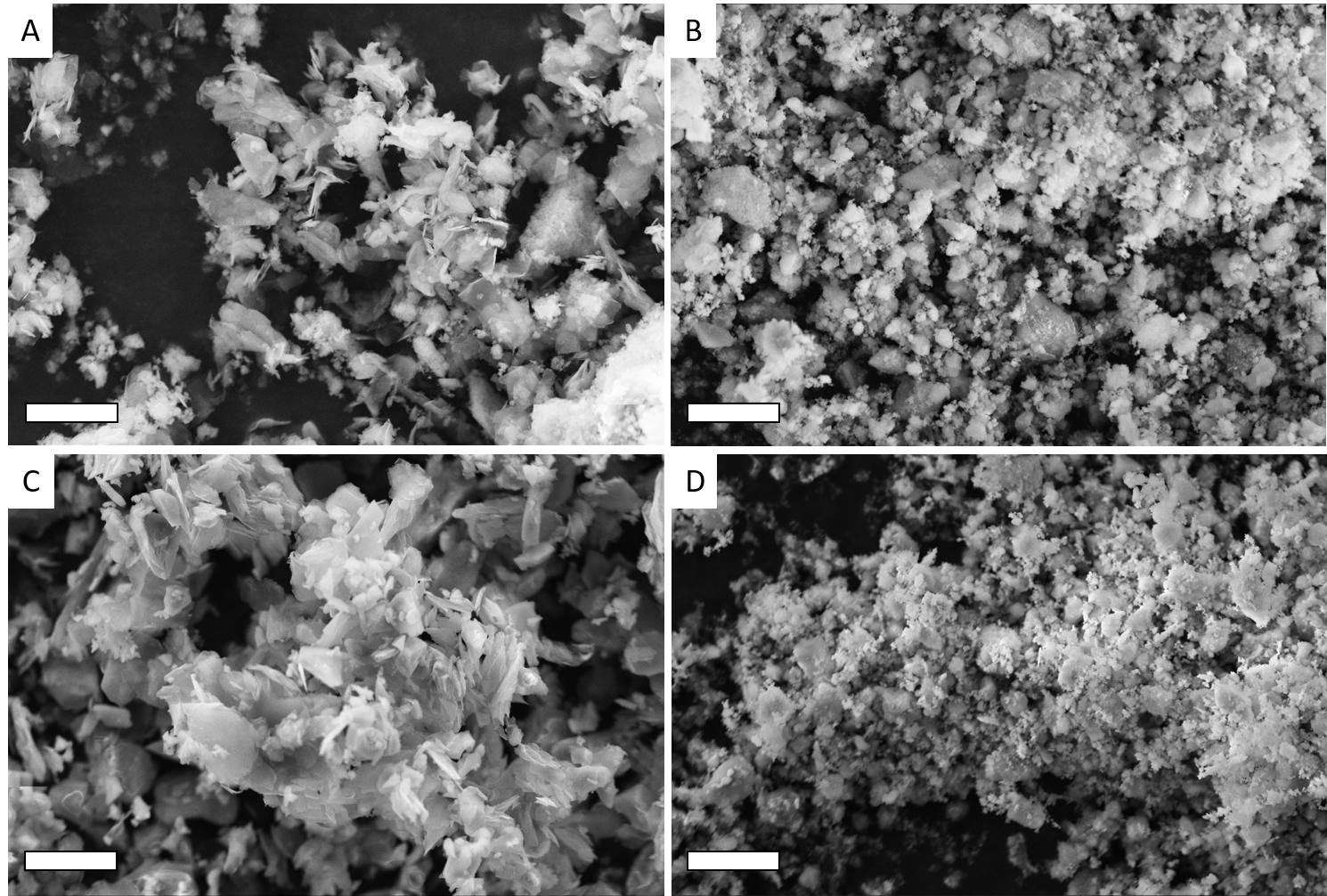}
\caption{Scanning electron microscope images of four mixtures. A) Forsterite-graphite 7:3, B) silica-forsterite 7:3 , C) spinel-graphite 7:3, and D) silica-magnetite 7:3. All the scale bars represent \SI{10}{\micro\metre}.}
    \label{fig:SEM_mix}
\end{figure*}

In Table \ref{tab:dust} we list the mineral powders that were used in the experiments. These include silicates (silica, forsterite, and fayalite), spinel-group minerals (magnetite, Mg-spinel) and graphite. The table also includes the reflectance $R$ of the pure powders measured at 530 nm using an integrating sphere for homogenous illumination. In general, the samples can be divided into two groups according to their reflectance: dark powders ($R<0.5$, magnetite, graphite, and fayalite) and bright powders ($R>0.5$, silica, forsterite, and Mg-spinel). Table \ref{tab:mixtures} lists the different mixtures with abbreviations measured in this work. The mass ratios of the two components mixed together (1:0, 9:1, 4:1, 7:3, 1:1, 1:3, and 0:1) are equivalent to 100-0\%, 90-10\%, 80-20\%, 70-30\%, 50-50\%, and 0-100\% of the total mass being first endmember and second endmember. The grain size ranges are comparable between different minerals; they are mainly about \SI{1}{\micro\metre}. We acquired scanning electon microscope (SEM) images of forsterite, spinel, and fayalite, confirming that these powders are mainly composed of \SI{}{\micro\metre} and sub-\SI{}{\micro\metre} sized particles. We acquired SEM images of the endmembers (Fig. \ref{fig:SEM}) and of some of the mixtures (Fig. \ref{fig:SEM_mix}). The grain shapes are generally irregular and within the size ranges provided by the supplier. Forsterite and magnetite mainly have sub-\SI{}{\micro\metre} grains, while graphite and fayalite have larger grains, although sub-\SI{}{\micro\metre} grains are still present and the surfaces of larger grains display features at the sub-\SI{}{\micro\metre} scale.

\begin{center}
\begin{table*}[ht]
        \centering
        \caption{List of minerals used in the experiments.}
        \label{tab:dust}
        \begin{tabular}{lcccc} 
                \hline
                Mineral & Ch. formula & Grain size & Reflectance (530 nm) & Supplier\\
                \hline
                Silica & \ce{SiO2}& $<$\SI{10}{\micro\metre}  & $0.650\pm0.013$ & Honeywell Fluka\\
                Magnetite & $\ce{Fe3O4}$ & $<$\SI{5}{\micro\metre}  & $0.010\pm0.001$ & Sigma Aldrich\\
                Graphite & \ce{C} & $<$\SI{20}{\micro\metre} & $0.028\pm0.002$ & Sigma Aldrich\\
                Forsterite & $\ce{Mg2SiO4}$ & $<$\SI{15}{\micro\metre}  & $0.758\pm0.014$ & F.J. Brodmann \& Co\\
                Spinel & $\ce{MgAl2O4}$ & $<$\SI{15}{\micro\metre} & $0.822\pm0.013$ & F.J. Brodmann \& Co\\
                Fayalite & $\ce{Fe2SiO4}$ & $<$\SI{15}{\micro\metre} & $0.158\pm0.004$ & F.J. Brodmann \& Co\\         
                \hline
        \end{tabular}
\end{table*}
\end{center}
\begin{center}
\begin{table*}[ht]
        \centering
        \caption{List of measured mixtures.}
        \label{tab:mixtures}
        \begin{tabular}{lllcc} 
                \hline
                Powder 1 & Powder 2 & Powder 3 & Mass ratio & Abbreviation\\
                \hline
                Silica & Graphite & $-$ & 99:1, 9:1, 4:1, 7:3, 1:3& si-graph\\
                Mg-spinel & Graphite & $-$ & 9:1, 4:1, 7:3, 1:3& spi-graph\\
            forsterite & Graphite & $-$ & 9:1, 4:1, 7:3, 1:3& fo-graph\\
                forsterite & Mg-spinel & $-$ & 9:1, 4:1, 7:3, 1:3& fo-spi\\
                Magnetite & Graphite & $-$ & 9:1, 4:1, 7:3, 1:3& mt-graph\\ 
                Forsterite & Fayalite & $-$ & 9:1, 4:1, 7:3, 1:3& fo-fa\\
                Forsterite & Silica & $-$ & 9:1, 4:1, 7:3, 1:3& fo-si\\
                Silica & Magnetite & $-$ & 9:1, 4:1, 7:3, 1:3& si-mt\\
                Silica & Forsterite & Graphite & 1:1:1, 2:9:9, 9:2:9, 9:9:2& si-fo-graph\\
                Silica & Magnetite & Graphite & 1:1:1, 2:9:9, 9:2:9, 9:9:2& si-spi-graph\\
                \hline
        \end{tabular}
\end{table*}
\end{center}

\subsection{Sample preparation}
In order to create binary mixtures, we weighed the two end members to the correct mass ratio. We subsequently mixed them until we obtained a homogeneous sample. 

As silica can easily create aggregates of cm size \citep{blum2006physics}, we used a binary mixture of silica and graphite to study the effect of aggregation on the negative polarization. After a homogeneous mixture was obtained, aggregates were created by gently moving the mixture in a bowl. Then we sieved the aggregates through a \SI{200}{\micro\metre} sieve to obtain fine aggregates. Larger aggregates are easily breakable if passed through a sieve, and therefore we individually chose aggregates larger than 2 mm that were then gently placed in the sample holder. Generally, the aggregates formed with this method reach sizes up to $\sim 1$cm.

The sample holder used for all our measurements consists of a plastic petri dish covered by black aluminum tape. The height of the sample is then adjusted in the enclosure so that the sample surface coincides with the rotation axis of the arm holding the incident light. In this way, the light spot always illuminates the center of the sample holder at the different phase angles. The sample holder is 5 mm deep and is filled with the sample in such a way that the walls and the bottom of the sample holder are completely concealed by the sample. Furthermore, the sample holder is covered by black aluminum tape to minimize the risk of polarization signal from the edges of the sample holder at large phase angles. The dust sample is gently deposited on the sample holder, without touching the surface to avoid compression of the powder. 

From the SEM images of the mixtures (Fig. \ref{fig:SEM_mix}), it is clear that the mixing procedure is effective at the particle level. Particles of different materials are well intermixed and adjacent to each other.

\subsection{Data acquisition}\label{data}
The polarized light reflected by the sample was measured with $\sim35$ different phase angles ranging from \ang{0.8} to \ang{75}. At angles smaller than \ang{30}, the sampling was smaller in order to better depict the behavior of the sample at small phase angles. 

Comparing the polarization of different samples, we consider different sources that contribute to the total error. Variations resulting from the sample geometry: each phase angle curve is the result of at least five measurements in which the sample azimuth was varied by \ang{45} incremental steps, essentially dampening the contribution of geometric effects related to nonflat samples.The presented phase curve average and standard deviation are taken over these repetitions. When the same sample is prepared different times and measured, the polarization signal is slightly different. This is due to the differences in the mixing process and in the sample preparation (i.e., geometrical effects are different for each sample). The differences between the average phase curves of the same sample mixed different times are minimal and give a standard deviation of $7\cdot10^{-5}$ (negligible error on the mixing). On the other hand, the standard deviation of each sample due to the azimuthal rotation is two orders of magnitude larger, depending on their different rough surfaces. This is the most important source of error when two different mixtures are compared.

From the repetition of the silica-graphite mixture (99:1 mass ratio), we estimated the maximum variation in polarization due to geometrical effects when preparing the sample to be $\pm0.03\%$. This error is to be used when the polarization of different mixtures is compared. The error on the evaluation of $\alpha_{min}$ is $\pm$\ang{0.5} because all curves are sampled with \ang{1} step around the polarization minimum. The error on the inversion angle was estimated using the maximum range of the geometrical error on the repetitions of the mixture, and it is $\pm$\ang{0.3}.

The reflectance of the dust powders (Table~\ref{tab:dust}) has been measured with a camera (CS126MU, Thorlabs) and through the use of an integrating sphere in combination with the 530 nm LED source for homogeneous illumination. In this configuration, the measured reflectance is a hemispherical-directional reflectance, but we refer to it as reflectance in this work. A spectralon target was used for image calibration. The errors associated with the measurements are the standard deviation of the pixel signal of the selected region of interest over the sample. 

\section{Results}

\begin{figure*}
\centering
\includegraphics{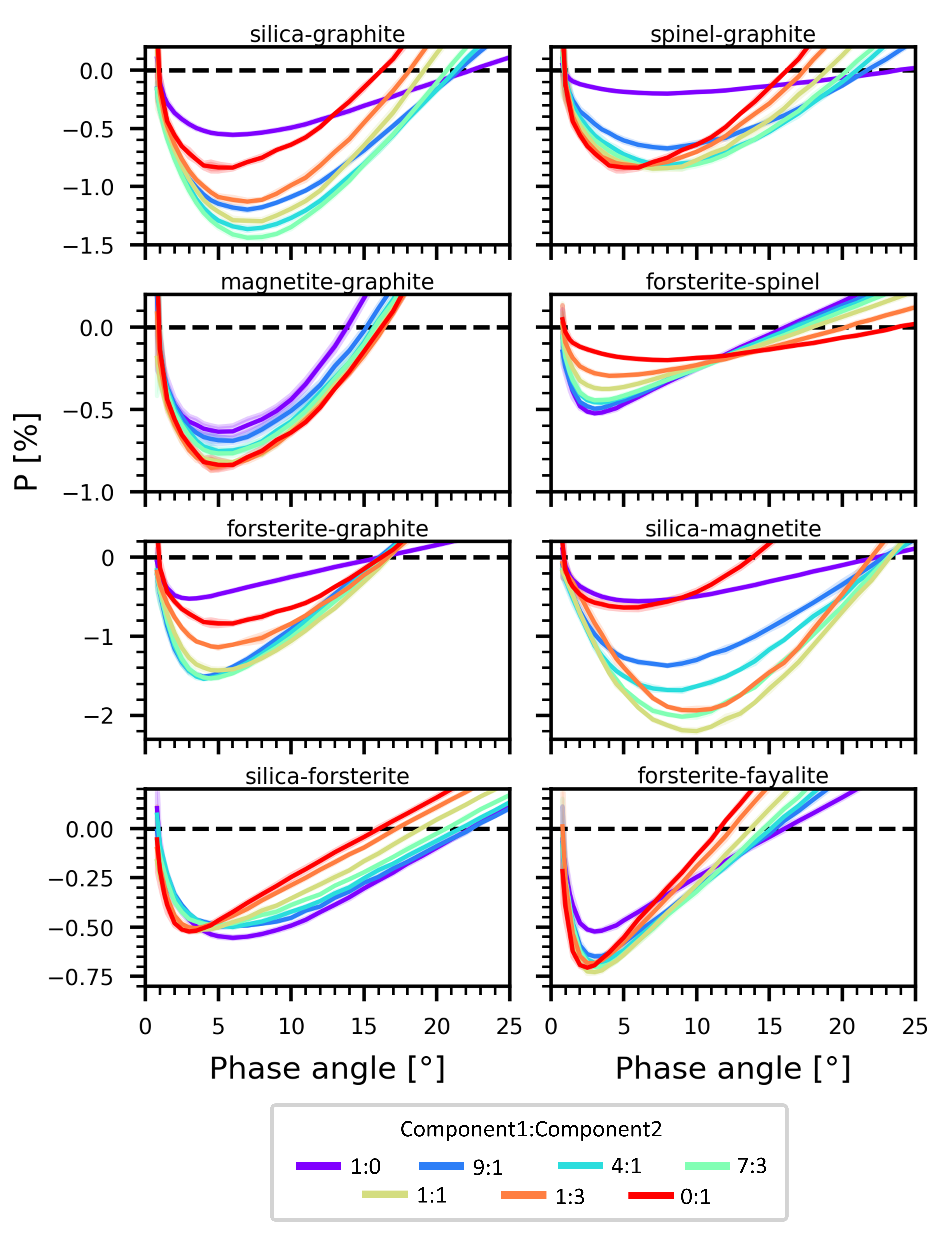}
\caption{Polarization phase curves of binary mixtures. The two mineral components are mixed with mass ratios shown in different colors. The endmembers of the mixture are plotted in purple and red. The mass ratios of the two components mixed together (1:0, 9:1, 4:1, 7:3, 1:1, 1:3, and 0:1) are equivalent to 100-0\%, 90-10\%, 80-20\%, 70-30\%, 50-50\%, 25-75\%, and 0-100\% of the total mass being first and second endmember. The shaded areas around the curves denote the standard deviation, which in this figure is often smaller than the line width.}
\label{fig:binary}
\end{figure*}

We measured the polarization of different mixtures at small phase angles to understand how the polarization minimum $|P_{min}|$, the phase angle at minimum of polarization $\alpha_{min}$ , and the inversion angle $\alpha_{inv}$ change with respect to the dust endmembers of the mixture (see also Fig. \ref{fig:binary}). In Appendix \ref{table:appendix} we present the main properties of the polarization phase curve of all the mixtures.

\subsection{Binary dust mixtures}
We prepared different mixtures and investigated the negative polarization depending on the mixing ratio of the two components (Fig. \ref{fig:binary}). The mixtures were silica-graphite, spinel-graphite, magnetite-graphite, forsterite-spinel, forsterite-graphite, silica-magnetite, silica-forsterite, and forsterite-fayalite (see also Table \ref{tab:mixtures}). The mixtures cover a range of contrast in reflectance between endmembers: dark-dark mixtures (magnetite-graphite), dark-bright mixtures (silica-graphite, spinel-graphite, forserite-graphite, silica-magnetite, and forsterite-fayalite), and bright-bright mixtures (forsterite-spinel and silica-forsterite). For most of the mixtures (spinel-graphite, magnetite-graphite, forsterite-spinel, silica-forsterite, and forsterite-fayalite), the phase functions of the different mixing ratios change monotonically between the phase functions of the pure endmembers. In three of the dark-bright mixtures, we observe a nonmonotonic behavior of the phase function of different mixing ratios. In particular, some mixing ratios of silica-magnetite, forsterite-graphite, and silica-magnetite show higher $|P_{min}|$, $\alpha_{min}$ and $\alpha_{inv}$ than the phase curves of the endmembers. 

Here we present a summary of our observations (for the mixture silica-graphite, we sieved the mixture to remove all aggregates larger than \SI{200}{\micro\metre}). In some of the mixtures, the mixture behaves very differenty from the two pure minerals that were mixed (called hereafter endmembers of the mixture). We observe a deepening of $|P_{min}|$ respect with both the endmembers in the case of silica-graphite, forsterite-graphite, silica-magnetite, while the other mixtures do not show deepening of $|P_{min}|$ (within error). Of the three, the maximum of $|P_{min}|$ is found for silica-magnetite, starting at about $P_{min}=-0.5\%$ and reaching $P_{min}=-2.2\%$ when the mass ratio of the two endmembers is 1:1.
The mass ratio at which $|P_{min}|$ is found varies with the different mixtures (e.g., 7:3 for silica-graphite and 4:1 for forsterite-graphite). 

Generally, the inversion angle of the mixtures varies monotonically from one of the endmembers $\alpha_{inv}$ to the other. There are two exceptions: forsterite-graphite and silica-magnetite. In both cases, some of their mixtures have a larger inversion angle than both endmembers (the endmembers with a larger inversion angle are forsterite and silica). In the forsterite-graphite mixture, the maximum inversion angle is reached in the 1:1 mixture with $\alpha_{inv}=16.7\pm$\ang{0.3} (to be compared with the inversion angle of forsterite, $\alpha_{inv}=16.0\pm$\ang{0.3}). In the case of silica-magnetite, the maximum inversion angle is reached by a 1:1 mixture with $\alpha_{inv}=23.2\pm$\ang{0.3} (to be compared with silica, $\alpha_{inv}=22.3\pm$\ang{0.3})

The binary mixtures that show a deepening of $|P_{min}|$ also show an increase in phase angle at which the minimum polarization occurs, $\alpha_{min}$. Silica-graphite, silica-magnetite, and forsterite-graphite mixtures all show a higher value of $\alpha_{min}$ than the endmembers $\alpha_{min}$. As in the case of $|P_{min}|$, the maximum $\alpha_{min}$ does not occur at fixed mass ratios, but depends on the minerals that are mixed together. The highest $\alpha_{min}$ compared to the endmembers is given by silica-magnetite, with $\alpha_{min}=10\pm$\ang{0.5} for the 1:1 mass ratio mixture (to be compared with $\alpha_{min}=6\pm$\ang{0.5} of pure silica).

The change in magnitude of $P_{min}$ with varying mixing ratios is generally nonlinear, in particular for the mixtures with a bright and a dark component. In the mixture spinel-graphite, for example, adding 10\% of graphite to spinel changes $P_{min}$ strongly, while adding more graphite results in only slight changes. Finally, if 25\% of spinel is added to graphite, the changes in phase curve are practically invisible without a precise polarization measurement.

\subsection{Ternary dust mixtures}
\begin{figure}
\centering
\includegraphics[width = \columnwidth]{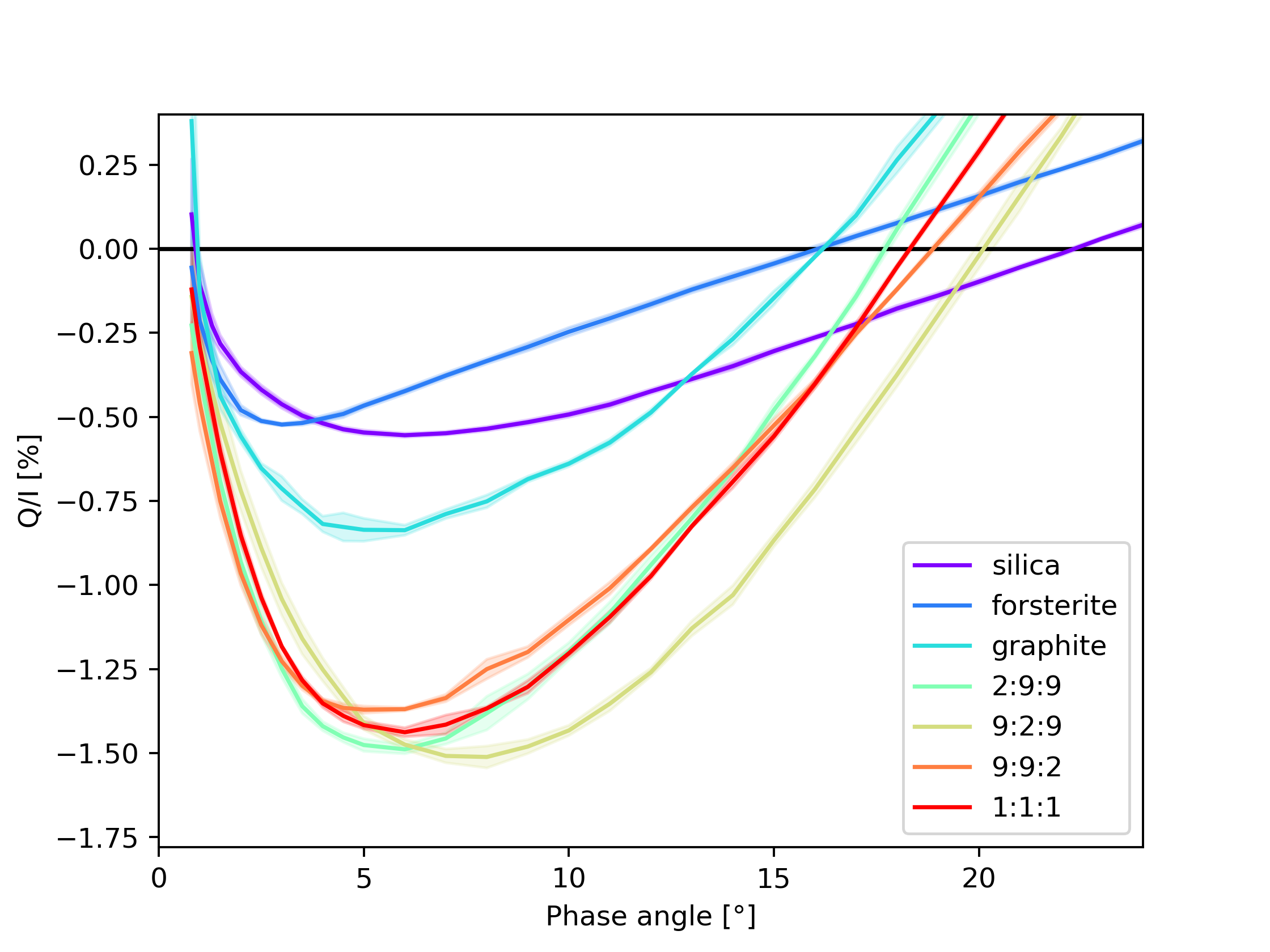}
\caption{Negative polarization of the ternary mixture silica-forsterite-graphite. The three mineral components are mixed, and the mass ratios are shown in different colors. The shaded areas denote the standard deviation.}
    \label{fig:ternary1}
\end{figure}

\begin{figure}
\centering
\includegraphics[width = \columnwidth]{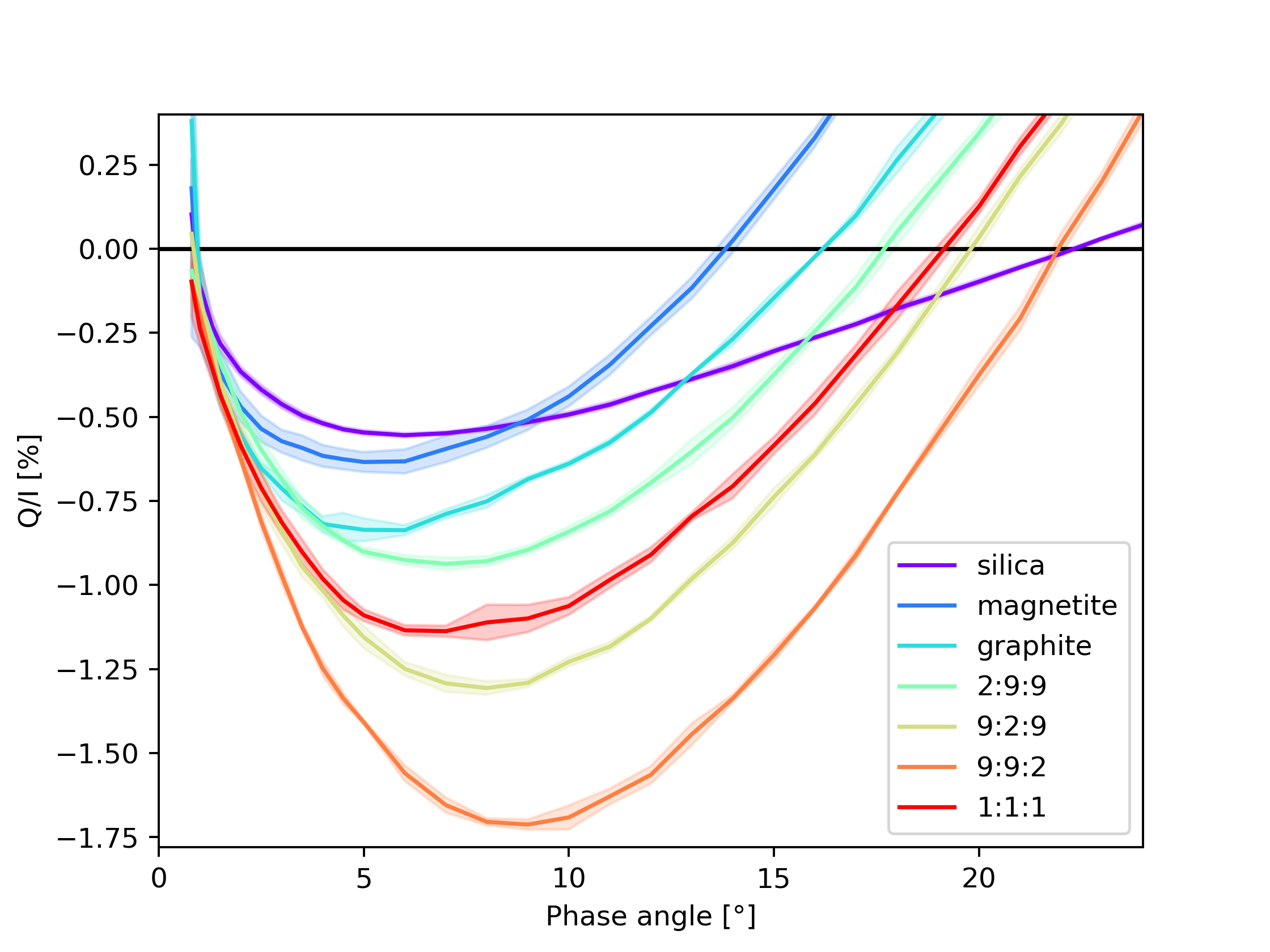}
\caption{Negative polarization of ternary mixture silica-magnetite-graphite. The three mineral components are mixed, and the mass ratios are shown in different colors. The shaded areas denote the standard deviation.}
    \label{fig:ternary2}
\end{figure}

Two ternary mixtures were prepared with silica, forsterite, and graphite (si-fo-graph) and silica, magnetite and graphite (si-mt-graph). The measured mixing ratios for the two ternary mixtures are 1:1:1, 2:9:9, 9:2:9, and 9:9:2 (corresponding to a weight percentage of the three minerals of $33.3-33.3-33.3\%$, $10-45-45\%$, $45-10-45\%$, and $45-45-10\%$). As expected, $|P_{min}|$ increases when bright and dark material are mixed together (Fig. ~\ref{fig:ternary1}-\ref{fig:ternary2}). The inversion angles of the ternary mixtures are within the two endmembers with lower and higher inversion angle, that is, forsterite-silica for the first ternary mixture, and magnetite-silica for the second one. Another interesting result is that the phase angle of the minimum polarization is larger for the mixtures than for the single endmembers, reaching $\alpha_{min}=8.0\pm$\ang{0.5} in the 9:2:9 si-fo-graph mixture and $\alpha_{min}=9.0\pm$\ang{0.5} in the 9:9:2 si-mt-graph mixture.

\subsection{Aggregates of silica-graphite}
\begin{figure}
\centering
\includegraphics[width = \columnwidth]{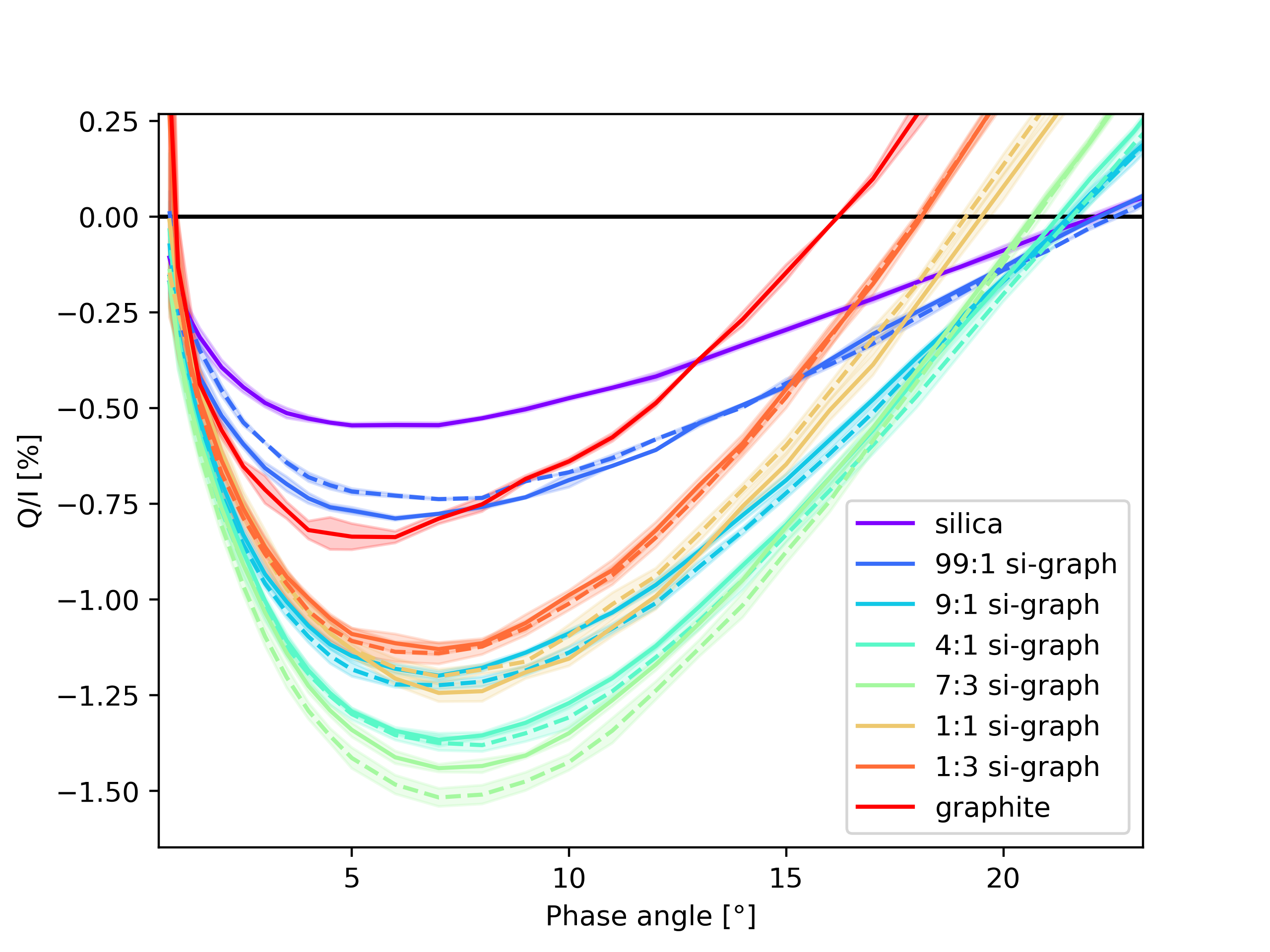}
\caption{Negative polarization of a binary mixture silica-graphite. Solid lines represent aggregate sizes smaller than \SI{200}{\micro\metre}, and dashed lines represent aggregate sizes larger than 2 mm. The shaded areas around the curves are the standard deviations of the measurements when the sample is rotated along the azimuthal axis.}
    \label{fig:aggregates}
\end{figure}

Silica can easily form cm size aggregates due to strong Van der Waals interactions between the particles \citep{blum2006physics}. We measured different amounts of graphite in silica (99:1, 9:1, 4:1, 7:3, 1:1, and 1:3) with two different aggregate sizes for each mixing ratio: aggregates smaller than \SI{200}{\micro\metre,} and aggregates larger than 2 mm (Fig.~\ref{fig:aggregates}). As we already observed for the aggregates that are smaller than \SI{200}{\micro\metre}, larger aggregates ($>2$ mm) follow the same evolution pattern when the graphite mixing ratio is increased. $|P_{min}|$ increases reaching $\simeq1.5$\% (7:3 mass ratio silica-graphite), and the phase angle at the polarization minimum increases up to $\alpha_{min}=7.5\pm$\ang{0.5}, as compared to $\alpha_{min}=6.0\pm$\ang{0.5} for both silica and graphite. While the inversion angle of most mixing ratios is between the inversion angles of pure silica and graphite, we observed that for the two aggregate sizes with a 99:1 silica-graphite mass ratio, the measured inversion angle exceeds the inversion angle of the pure endmembers. However, the difference was within the error range estimated for comparing different mixtures ($\alpha_{inv}=22.6\pm$\ang{0.3} to the inversion angle of silica $\alpha_{inv}=22.1\pm$\ang{0.3}). This behavior, in which the inversion angle exceeds that of the pure endmembers, was also observed for the other binary mixtures that show a deepening of $|P_{min}|$ (silica-magnetite and forsterite-graphite).

Generally, the differences between the two aggregate sizes are very small. Most of them fall within the error on the polarization when different mixtures (and geometries) are compared. For most of the mixtures, the larger aggregate size shows a slightly lower $P_{min}$.

\section{Discussion}
We presented the polarization phase-angle dependence of different mineral powders, their binary and ternary mixtures, and the effects of different aggregate sizes. Our results depict a complicated but interesting picture of phase function behavior that can be used to interpret the polarimetric properties of asteroids. 

\subsection{Mixing effect}
Our results indicate that mixing different mineral powders can cause a surge of $P_{min}$. It is possible to relate this effect to the mixing of different mineralogies, and exclude the influence of other parameters (grain shapes, grain sizes, and porosity).

All the analyzed endmembers show very irregular particles with sharp edges and different morphologies (e.g., flat for graphite, rounder for magnetite). Because the shape is so various and the polarization increases in mixtures with very different particle shapes, this parameter is apparently not responsible for the surge in negative polarization upon mixing.

Our samples also show different size distributions within the maximum grain size provided by the supplier. Forsterite and magnetite are mainly composed by small, sub-\SI{}{\micro\metre} sized particles, while the grains of graphite and fayalite are closer to the $5-$\SI{10}{\micro\metre} average size (Fig. \ref{fig:SEM}). Interestingly, mixtures can have similar grains sizes and completely different phase curves. For example, the silica-forsterite 7:3 mixture and the silica-magnetite 7:3 mixture show no obvious differences in grain sizes (Fig. \ref{fig:SEM_mix} B and D), but silica-magnetite shows an impressive deepening in polarization minimum upon mixing, while silica-forsterite does not. This also holds for forsterite-graphite and spinel-graphite, which both have similar particles sizes. Only the first mixture shows a deepening of the negative polarization, however. Furthermore, the presence of sub-\SI{}{\micro\metre} sized particles in one endmember is not a sufficient condition to cause the deepening of the negative polarization when mixed with another endmember. As an example, in the case of the magnetite-graphite mixture, no deepening is observed, while graphite-forsterite mixtures show a deepening in $P_{min}$ . These results seem to exclude that the deepening of the negative polarization originates mainly in the particle size distribution of the mixtures.

Finally, we considered the porosity of the sample after mixing as a possible source of the observed effect. \citet{shkuratov_opposition_2002} experimentally investigated the effect of the porosity of a granular material on the shape of its polarization phase curve. They found that the minimum of polarization can deepen and shift to larger phase angles, although the analysis was limited to two very bright powders, namely MgO and SiO$_{2}$. Other theoretical works (e.g., \citet{mishchenko2009direct}) indeed showed that the dust-packing density can shape the negative polarization minimum and the inversion angle. Although the samples are deposited in the sample holder without compressing their surface, the mixing of different mineral species could still change the packing density of the sample. To investigate the effect of porosity on our samples, we present in Appendix \ref{compression} the effect of a compression experiment on a mixture and its two end members, namely silica, magnetite and their mixture 1:1 (Fig. \ref{fig:compression}). Compressing the samples with a pressure of 1100 kg m$^{-2}$ increased the minimum of polarization by approximately 0.2-0.4\% in the case of silica and the mixture 1:1, but increased it by about 0.2\% in the case of magnetite. The inversion angle does not show a consistent behavior upon compression either: it increases for silica, is almost the same for the silica-magnetite mixture, and decreases for compressed magnetite. In general, the decrease in $P_{min}$ upon mixing cannot be explained by a difference in the porosity state of the sample. While compression effects are not the purpose of this work, we note that a full laboratory investigation on natural samples is lacking in the literature, and our results show that the compression of different mineral powders can change the shape of the negative polarization in opposite ways (e.g., silica vs magnetite). Future laboratory work on this topic will be an important extension of this manuscript.

We conclude that although the negative polarization of our samples is determined by the overall porosity, particle size, and particle shape, none of these parameters controls the surge in negative polarization upon mixing. In Sect. \ref{sec:reflectance} we discuss the possibility that the extent of negative polarization of a mixture is given by the photometric contrast between the two endmembers.

\subsection{Binary mixtures}

\begin{figure*}[ht!]
    \centering
        \includegraphics[width=425pt]{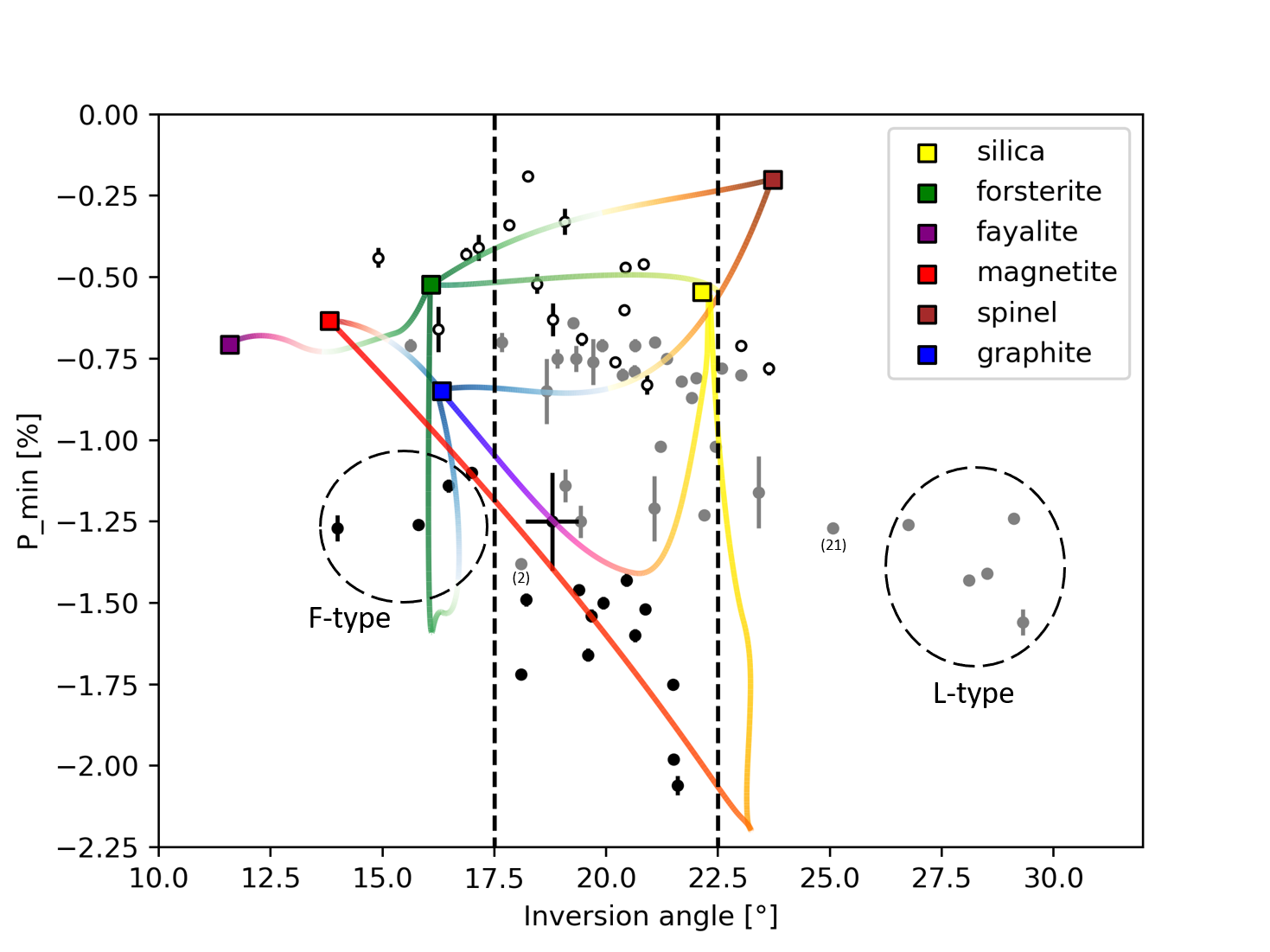}
    \caption{$P_{min}$ vs $\alpha_{inv}$ for asteroids and laboratory measurements. The polarimetric data of asteroids are taken from \citet{belskaya_f-type_2005}, \citet{cellino_2016}, and \citet{belskaya2017refining}. Low-, moderate-, and high-albedo asteroids are plotted with dark, gray, and white dots, respectively. The endmembers of our binary mixtures are colored squares, and the binary mixture data fitted with a spline are represented by a line connecting two endmembers. The colors of the lines do not represent the mixing ratios of the endmembers, but have only an illustrative purpose. The group of L-type asteroids (Barbarians), F-type asteroids, (2) Pallas, and (21) Lutetia is highlighted}
    \label{fig:painv}
\end{figure*}

It has been demonstrated that asteroids cluster together in the $P_{min}-\alpha_{inv}$ space according to their class types \citep{belskaya2017refining}. Generally, this indicates that asteroids in the same family share similar mineralogical compositions and physical properties. To some extent, this is also affected by the asteroid albedo. Asteroids with a high albedo in the V band (more than 0.2) have a higher $P_{min}$ than moderate-albedo asteroids (0.1-0.2 in V), and the darkest asteroids (C, Ch, and B classes, with an albedo lower than 0.1) populate the lowest $P_{min}$ region. In this context, \citet{cellino_2016} noted that there is some degree of mixing between moderate- and low-albedo asteroids defined in the region $P_{min}=-1.1\%-1.4\%$ and $\alpha_{inv}=18-$\ang{21} in Fig.~\ref{fig:painv}. \citet{dollfus1989photopolarimetry} interpreted the fact that terrestrial rocks and lunar fines show smaller and larger polarimetric inversion angles, respectively, than most asteroids as an indication that the surface of asteroids contains coarser material than the surface of the Moon. More recently, two new classes of rare asteroids have been identified \citep{belskaya_f-type_2005,cellino2006strange}: the F-class asteroids, which show small inversion angles ($14-$\ang{16}), and L-type asteroids (''Barbarians``), with inversion angles larger than \ang{25}. These asteroid classes are outliers to the normal distribution of asteroids in the $P_{min}-\alpha_{inv}$ space.

In Fig.~\ref{fig:painv} we present the results of our binary mixtures and their coverage within the $P_{min}-\alpha_{min}$ space. Each line connecting the endmembers was obtained by fitting a spline through the data points of $P_{min}$ and $\alpha_{inv}$ versus the concentration of one of the endmembers. 

The minerals used here are not distributed in exactly the same regions as described by \citet{dollfus1989photopolarimetry} for terrestrial rocks. In particular, both silica and spinel show larer inversion angles than other samples. 
Mixing is an efficient way of exploring the $P_{min}-\alpha_{inv}$ variable space. Moreover, because two endmembers with similar size distribution are mixed, we expect that this result is free from grain size effects (the mixture size distribution is similar to that of the two endmembers). This means that $P_{min}$ and $\alpha_{inv}$ are indeed related to the mineralogy of the sample.  
Although the endmembers are positioned in the top half of the $P_{min}-\alpha_{inv}$ space, mixing of bright and dark minerals allows the exploration of lower $P_{min}$ values, and in some cases, also higher values of the inversion angle (e.g., as observed for silica-magnetite).

The presence of several different minerals (and organics) clearly affects the overall signal of asteroids (see ~\ref{sec:ternary_mixtures}). In addition, we did not consider other parameters that would influence the polarization signal, such as grain size and regolith porosity. However, we would like to underline that observational data and theoretical results on grain sizes, mineralogy, and surface properties of the regolith coupled to experimentally derived polarimetric phase curves could be a powerful tool for constraining these variables.

\subsubsection{F-type, L-type, and other asteroids}
F-type asteroids have been proposed to be covered by a homogeneous dark regolith that could explain their small inversion angles and relatively small $P_{min}$ as compared to C-type asteroids \citep{belskaya_f-type_2005}. We explored the $P_{min}-\alpha_{inv}$ space in which F-type asteroids reside through a mixture of graphite and forsterite. While this particular composition is certainly not relevant for this class of objects, we demonstrate that this region can efficiently be explored with mixtures of bright and dark material. The mixture that is closest to the F-type albedo, $P_{min}$ and $\alpha_{inv}$, is a 1:3 forsterite-graphite mixture ($R\sim0.03\pm0.01$,$P_{min}=-1.14\pm0.03\%$, and $\alpha_{inv}=16.6\pm0.3$). The polarization properties of F-type asteroids are thus still compatible with the mixing of dark material with small parts of bright minerals, and no homogeneity of the surface must be invoked. 

Barbarians are positioned in a region in $P_{min}-\alpha_{inv}$ space that is not explored by our binary mixtures. Their polarimetric properties have been associated with spinel-bearing CAIs on the surface that have a high refractive index and consequently a large inversion angle \citep{burbine1992s,sunshine_ancient_2008,devogele_new_2018}.
Our Mg-spinel sample has a smaller inversion angle than the Barbarians. When mixed with graphite, it shows no deepening in $P_{min}$, and when only 10\% of graphite are added, the inversion angle decreases from $\alpha_{inv}=23.7\pm$\ang{0.3} to $\alpha_{inv}=21.7\pm$\ang{0.3}. When spinel is mixed with graphite with a 1:3 mass ratio, the signature polarizance by spinel is completely hidden, and the total resembling pure graphite. 

The polarization minima of Barbarians range from -1.2 to -1.6\%, similar to C and B-type asteroids, but Barbarians also display a higher albedo (0.15-0.2 in V band). A plausible explanation for these values is the mixing of bright (CAIs) and dark material. Furthermore, another non-Barbarian asteroid, (21) Lutetia, which shares a large inversion angle ($\alpha_{inv}=$\ang{25}) with Barbarians, is thought to be rich in very fine regolith \citep{keihm2012interpretation}.

Another distinct phenomenon that we observe with our binary mixtures is the increase in inversion angle with increasing contrast of the two endmembers (pure Mg-spinel excluded). Other studies found similar results for mixing bright and dark materials: \citet{zellner1977asteroid} found that the inversion angle increases by \ang{3} compared to pure fine silicates when 10\% of 10 nm soot were added, and \citet{Shkuratov1987} found that a 1:1 mixture of submicron MgO and Fe$_{2}$O$_{3}$ shows a \ang{9} larger inversion angle with respect to the largest inversion angle of the endmembers (MgO). In our sample, the largest increase in inversion angle is given by a 1:1 mixture of silica and magnetite, with an excess of \ang{1} compared to the inversion angle of pure silica. We expect that mixtures of very fine dark and bright particles could increase the inversion angle even more substantially compared with the single endmembers. 

We therefore propose that the polarimetric behavior of Barbarians is not merely related to the presence of a single mineral, but to a combination of very fine regolith that contains both bright and dark components that are mixed finely together. In-depth polarimetric measurement of CAIs and dark matrix from meteorites are necessary to make further progress on this question.

In addition, mixtures
of bright and dark components can also explain that in the same $P_{min}-\alpha_{inv}$ region  ($P_{min}=-1.1\%-1.4\%$ and $\alpha_{inv}=18-$\ang{21}), a certain mixing occurs between low and moderate-albedo asteroids. The asteroid (2) Pallas is a good example. While it is classified as a B-type asteroid, it has a higher albedo (0.145) than the other B-class asteroids. The reason for the higher albedo has been suggested to be the presence of salts on the surface \citep{marsset2020violent}. Even if these salt spots are localized on its surface, it is plausible that the salt is mixed to some degree with the asteroid dust. This could give rise to the polarization mixing effect and thus explain why its $P_{min}$ remains low even when albedo is higher than that of asteroids of the same class. 
Other M-type asteroids that have a quite low $P_{min}$ but moderate albedo ($P_{min}=-1.1\%-1.4\%$ and $\alpha_{inv}=18-$\ang{21}) can have some degree of mixing on their surface, which determines their polarimetric properties. Recent observations by \citet{belskaya2022polarimetry} showed that the negative branch of M-class asteroids can be correlated to different compositions of their surface. It might be divided into sub-groups depending on their different mineral compositions (represented by different meteorite analogs, such as irons and stony irons, and enstatite and high-iron carbonaceous chondrites).

\subsection{More complex mixtures} \label{sec:ternary_mixtures}
The mineralogical complexity of asteroids is much broader than simple binary mixtures. Our results with ternary mixtures show that a more complicate mineralogical mixture can result in different polarization phase curves. The phase curve does not only depend on the overall dark and bright materials. For instance, in the silica-forsterite-graphite mixture (Fig.\ref{fig:ternary1}), the 1:9:9 and 9:1:9 mixtures have the same amount of graphite, but different amounts of silica and forsterite (10-45\% of the sample mass in the first and 45-10\% in the second mixture). More silica as the bright component results in an increase in inversion angle and phase angle of the polarization minimum at more or less similar $P_{min}$. In the silica-magnetite-graphite mixture (Fig. ~\ref{fig:ternary2}), the 9:1:9 and 9:9:1 have the same amount of dark and bright material, while the first has only 10\% of magnetite and 45\% of graphite, and the second has 45\% of magnetite and 10\% of graphite. The difference of the two mixtures in terms of polarimetric properties is significant: $P_{min}$ decreases by 0.5\%, the inversion angle increases by \ang{2} and $\alpha_{min}$ increases by \ang{1}. 

Similar polarimetric experiments in the laboratory combined with astronomical observations could provide many important constraints on the mineralogical and physical properties of the regolith. When the main mineral constituents that contribute to the bright and dark components of asteroid regolith are known (deduced from spectroscopy, or from the associated class of meteorites), the polarization phase curve can provide great insight into the mixing ratios, grain sizes, and porosity.

\subsection{Aggregates}
We used different mixing ratios of graphite and silica with two different aggregate sizes in order to investigate the dependence of polarization on aggregate size at small phase angles. We find that the aggregate size does not play a significant role in changing the negative polarization, at least for aggregates up to cm size. It is possible that the negative polarization might vary on those asteroids where regolith is composed of more compact aggregates of very fine material. The compaction of fine powder increases $|P_{min}|$ and can in general change the shape of the negative polarization \citep{shkuratov_opposition_2002}. For asteroids on which very fine, porous regolith is expected, the phase function at small phase angles is not sensitive to possible aggregation processes, at least up to cm size aggregates. Above this limit, compaction could play an important role in shaping the negative polarization.

\subsection{Reflectance contrast between endmembers} \label{sec:reflectance}
One of the parameters we used to evaluate the photometric homogeneity of a granular material is the contrast parameter $K$, which can be defined as
\begin{equation}
    K=\frac{A_{l}-A_{d}}{A_{l}+A_{d}},
\end{equation}
where $A_{l}$ and $A_{d}$ are the albedos of the light and dark components of a mixture, respectively. Similarly, using our reflectance data, we can calculate a contrast parameter $K_{R}$. In the past, an increase in contrast parameter for different wavelengths has been correlated to an increase in $|P_{min}|$ and $\alpha_{inv}$ \citep{Shkuratov1987}. The contrast parameters for the silica-graphite, forsterite-graphite, and silica-magnetite are $K_{R}=0.92$, $0.93,$ and $0.97$ respectively. While this increase correlates with the maximum decrease of $P_{min}$ of the three mixtures ( $-1.44$, $-1.53$, $-2.20$\%, respectively), we observe no deepening in $P_{min}$ in the mixture spinel-graphite, even though the contrast parameter is very high: $K_{R}=0.93$. This indicates that the contrast between the two components alone does not determine the extent of the negative polarization of the mixture. Future experiments should aim to address this point and investigate the underlying causes for the increase in $|P_{min}|$ and $\alpha_{inv}$, and their relation to the mineralogy of the mixtures. 

\section{Conclusion}
We have consistently investigated the influence of mixing different minerals on their polarization phase function at small phase angles. We found that the polarization minimum, the inversion angle, and the phase angle of the minimum polarization are very sensitive to the mixing of bright and dark components. Furthermore, we observe that larer inversion angles and minimum phase angles can be reached by mixing different minerals, without changing the grain size distribution. More complex mixtures of minerals show different negative polarization properties of the endmembers, and aggregates up to cm sizes do not affect the negative polarization. Furthermore, the mixing effect dominates the negative polarization contribution over other parameters (particle size distributions, porosity, particles shape, and albedo of the mixed minerals).
We propose that this effect contributes to the polarization properties of particular classes of objects (F- and L-type asteroids) and other asteroids with unusual polarimetric features (mixing of low- and moderate-albedo asteroids in $P_{min}-\alpha_{inv}$ space). 

A good synergy between modeling, observations, and laboratory experiments has the potential of strongly aiding in interpreting the surface properties of regolith when reflected polarized light is observed. Future sample-return missions and in-situ highly sensitive polarimetric observations will greatly improve our understanding of asteroid regolith properties. This will help to interpret astronomical measurements and constrain laboratory simulations to more realistic mineralogies.

\begin{acknowledgements}\\
This work has been carried out within the framework of the National Centre of Competence in Research (NCCR), PlanetS, supported by the Swiss National Science Foundation (SNSF). We thank the anonymous reviewer for the useful insights which helped improving the manuscript.
\end{acknowledgements}

\bibliography{paper.bib}
\bibliographystyle{aa}

\begin{appendix} 

\section{Polarimetric phase curve data}
\centering          
\onecolumn
{
{\small\tabcolsep=3pt
\setlength{\LTleft}{2cm plus -1fill}
\setlength{\LTright}{\LTleft}
\begin{longtable}{lcccccc}  
\caption{Polarization phase curve data}\\
\hline
Minerals &Mass ratio & Aggregate size [mm]& $P_{min}$ ($\pm0.3$) \% & $\alpha_{inv}$ ($\pm$\ang{0.3}) & $\alpha_{min}$ ($\pm$\ang{0.5}) & $R$ [\%]\\ 
\hline
   silica (si) & $-$ & $<0.2$ & $-0.56$ & $22.2$ & $6.0$ & $65.0\pm1.0$\\
   Mg-spinel (spi) & $-$ & $-$ & $-0.20$ & $23.7$ & $8.0$ & $82.3\pm1.3$\\  
   graphite (graph) & $-$ & $-$ & $-0.84$ & $16.2$ & $6.0$ & $2.8\pm0.2$\\
   magnetite (mt) & $-$ & $-$ & $-0.63$ & $13.8$ & $5.5$ & $1.0\pm0.1$\\
   forsterite (fo) & $-$ & $-$ & $-0.52$ & $16.1$ & $3$ & $75.8\pm1.4$\\
   fayalite (fa) & $-$ & $-$ & $-0.71$ & $11.6$ & $2.5$ & $15.8\pm0.4$\\
   si-graph & $99:1$ & $<0.2$ & $-0.78$ & $22.2$ & $6.0$ & $35.1\pm4.0$\\
   si-graph & $99:1$ & $>2$ & $-0.74$ & $22.6$ & $7.0$ & $-$\\
   si-graph & $9:1$ & $<0.2$ & $-1.20$ & $21.5$ & $7.0$ & $15.0\pm0.7$\\
   si-graph & $9:1$ & $>2$ & $-1.22$ & $21.6$ & $7.0$ & $-$\\
   si-graph & $4:1$ & $<0.2$ & $-1.37$ & $21.6$ & $7.0$ & $10.5\pm1.6$\\
   si-graph & $4:1$ & $>2$ & $-1.38$ & $21.3$ & $8.0$ & $-$\\
   si-graph & $7:3$ & $<0.2$ & $-1.44$ & $20.7$ & $7.0$ & $9.3\pm0.5$\\
   si-graph & $7:3$ & $>2$ & $-1.52$ & $20.7$ & $7.0$ & $-$\\
   si-graph & $1:1$ & $<0.2$ & $-1.24$ & $19.5$ & $7.0$ & $4.3\pm0.6$\\
   si-graph & $1:1$ & $>2$ & $-1.20$ & $19.1$ & $7.0$ & $-$\\
   si-graph & $1:3$ & $<0.2$ & $-1.13$ & $18.1$ & $7.0$ & $3.7\pm0.4$\\
   si-graph & $1:3$ & $>2$ & $-1.14$ & $18.1$ & $7.0$ & $-$\\
   fo-spi & $9:1$ & $-$ & $-0.5$ & $16.5$ & $3.0$ & $77.4\pm3.0$\\
   fo-spi & $4:1$ & $-$ & $-0.46$ & $16.7$ & $3.5$ & $79.9\pm2.6$\\
   fo-spi & $7:3$ & $-$ & $-0.44$ & $17.3$ & $3.0$ & $71.7\pm2.6$\\
   fo-spi & $1:1$ & $-$ & $-0.37$ & $17.9$ & $3.5$ & $73.0\pm0.1$\\
   fo-spi & $1:3$ & $-$ & $-0.29$ & $20.2$ & $4.0$ & $69.9\pm2.8$\\
   fo-graph & $9:1$ & $-$ & $-1.51$ & $16.0$ &$4.0$ & $10.1\pm0.7$\\
   fo-graph & $4:1$ & $-$ & $-1.54$ & $16.2$ & $4.5$ & $7.6\pm0.8$\\
   fo-graph & $7:3$ & $-$ & $-1.53$ & $16.4$ & $4.5$ & $5.7\pm0.7$\\
   fo-graph & $1:1$ & $-$ & $-1.43$ & $16.7$ & $5.0$ & $3.7\pm0.5$\\
   fo-graph & $1:3$ & $-$ & $-1.14$ & $16.6$ & $5.0$ & $2.8\pm0.5$\\
   fo-fa & $9:1$ & $-$ & $-0.65$ & $15.4$ & $3.0$ & $45.4\pm1.4$\\
   fo-fa & $4:1$ & $-$ & $-0.68$ & $15.1$ & $3.5$ & $38.7\pm1.5$\\
   fo-fa & $7:3$ & $-$ & $-0.70$ & $14.6$ & $3.0$ & $32.3\pm1.1$\\
   fo-fa & $1:1$ & $-$ & $-0.73$ & $13.7$ & $3.0$ & $22.8\pm1.2$\\
   fo-fa & $1:3$ & $-$ & $-0.68$ & $12.4$ & $2.5$ & $17.5\pm0.5$\\
   si-mt & $9:1$ & $-$ & $-1.31$ & $22.8$ & $8.0$ & $13.9\pm1.4$\\
   si-mt & $4:1$ & $-$ & $-1.68$ & $23.2$ & $9.5$ & $8.6\pm1.5$\\
   si-mt & $7:3$ & $-$ & $-2.02$ & $23.1$ & $9.0$ & $5.6\pm0.6$\\
   si-mt & $1:1$ & $-$ & $-2.20$ & $23.2$ & $10.0$ & $3.9\pm0.6$\\
   si-mt & $1:3$ & $-$ & $-1.94$ & $22.0$ & $10.0$ & $1.9\pm0.4$\\
   si-fo & $9:1$ & $-$ & $-0.50$ & $22.4$ & $5.0$ & $61.4\pm3.1$\\
   si-fo & $4:1$ & $-$ & $-0.50$ & $21.6$ & $6.5$ & $66.4\pm2.8$\\
   si-fo & $7:3$ & $-$ & $-0.49$ & $20.7$ & $5.0$ & $64.8\pm2.5$\\
   si-fo & $1:1$ & $-$ & $-0.51$ & $18.9$ & $3.5$ & $76.5\pm1.4$\\
   si-fo & $1:3$ & $-$ & $-0.51$ & $17.3$ & $3.5$ & $76.1\pm1.6$\\
   mt-graph & $9:1$ & $-$ & $-0.69$ & $15.2$ & $6.0$ & $1.8\pm0.3$\\
   mt-graph & $4:1$ & $-$ & $-0.75$ & $15.7$ & $5.0$ & $1.9\pm0.3$\\
   mt-graph & $7:3$ & $-$ & $-0.76$ & $15.7$ & $5.0$ & $2.3\pm0.5$\\
   mt-graph & $1:1$ & $-$ & $-0.82$ & $16.1$ & $6.0$ & $2.2\pm0.1$\\
   mt-graph & $1:3$ & $-$ & $-0.85$ & $16.4$ & $4.5$ & $2.3\pm0.2$\\
   spi-graph & $9:1$ & $-$ & $-0.67$ & $21.7$ & $8.0$ & $10.3\pm0.7$\\
   spi-graph & $4:1$ & $-$ & $-0.82$ & $21.0$ & $8.0$ & $6.1\pm0.8$\\
   spi-graph & $7:3$ & $-$ & $-0.84$ & $20.3$ & $7.0$ & $5.9\pm0.8$\\
   spi-graph & $1:1$ & $-$ & $-0.85$ & $18.9$ & $7.0$ & $5.0\pm0.3$\\
   spi-graph & $1:3$ & $-$ & $-0.82$ & $17.4$ & $6.5$ & $2.4\pm1.7$\\
   si-mt-graph & $1:1:1$ & $-$ & $-1.14$ & $19.2$ & $7.0$ & $3.1\pm0.7$\\
   si-mt-graph & $2:9:9$ & $-$ & $-0.94$ & $17.7$ & $7.0$ & $2.5\pm0.4$\\
   si-mt-graph & $9:2:9$ & $-$ & $-1.31$ & $19.8$ & $8.0$ & $2.4\pm0.8$\\
   si-mt-graph & $9:9:2$ & $-$ & $-1.71$ & $22.0$ & $9.0$ & $3.9\pm0.8$\\
   si-fo-graph & $1:1:1$ & $-$ & $-1.44$ & $18.3$ & $6.0$ & $5.8\pm1.0$\\
   si-fo-graph & $2:9:9$ & $-$ & $-1.49$ & $17.7$ & $6.0$ & $4.6\pm0.8$\\
   si-fo-graph & $9:2:9$ & $-$ & $-1.51$ & $18.9$ & $8.0$ & $4.6\pm0.6$\\
   si-fo-graph & $9:9:2$ & $-$ & $-1.37$ & $22.0$ & $5.0$ & $12.7\pm2.1$\\   
   \\
\hline
\label{table:appendix} 
\end{longtable}
}}
\newpage
\twocolumn
{
\section{Polarimetric phase curves of compressed samples} \label{compression}
We present here the negative polarization of two endmembers (silica and magnetite) and their mixture 1:1 (Fig. \ref{fig:compression}) after compression. The samples were compressed with a hydraulic press up to a pressure of 1100 kg m$^{-2}$, producing a flat compacted surface. Further compression was not possible due to the fragility of the sample holder. Interestingly, the compression changed the negative polarization curves differently, depending on the material, with higher $|P_{min}|$ for silica and si-mt mixture, and lower $|P_{min}|$ for magnetite. The inversion angle after compression is larger for silica, equal for the si-mt mixture, and smaller for magnetite than in the uncompressed samples.
\begin{figure}[!htb]
\centering
\includegraphics[width = \columnwidth]{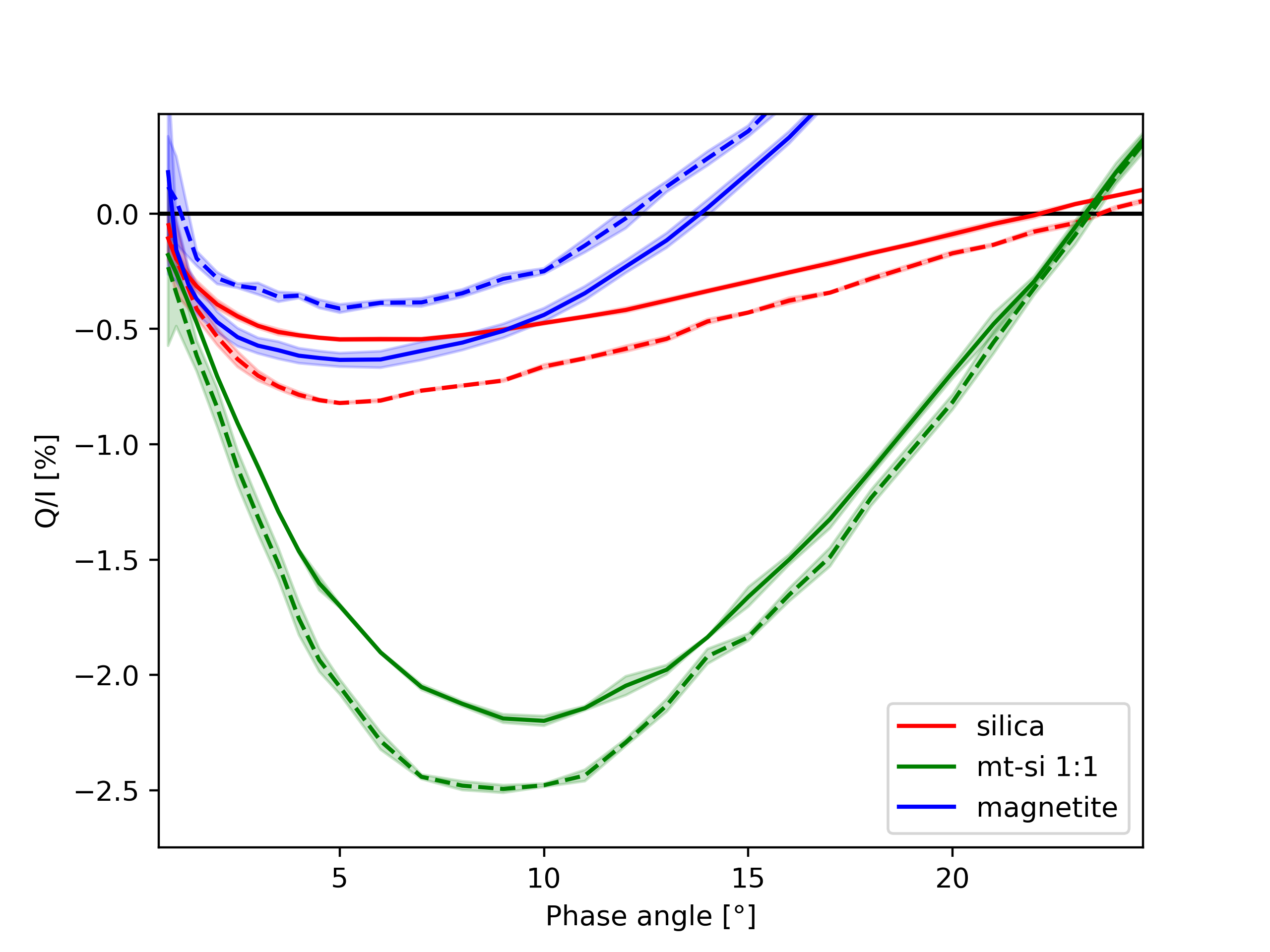}
\caption{Negative polarization of silica, magnetite, and their mixture 1:1 (solid line). The same samples have been compressed and measured again (dashed line).}
    \label{fig:compression}
\end{figure}
}
\end{appendix}

\end{document}